# Multi-sensorial interaction with a nano-scale phenomenon: the force curve

Sylvain Marliere[1,2], Daniela Urma[1], Jean-Loup Florens[1], Florence Marchi[2]

[1] ICA-ACROE, 46 Av. Félix Viallet, 38031 Grenoble Cedex, France
[2] LEPES, CNRS, 25 Avenue des Martyrs, 38042 Grenoble Cedex9, France
E-mail : marchi@grenoble.cnrs.fr

**Abstract.** Using Atomic Force Microscopes (AFM) to manipulate nano-objects is an actual challenge for surface scientists. Basic haptic interfaces between the AFM and experimentalists have already been implemented. The multi-sensory renderings (seeing, hearing and feeling) studied from a cognitive point of view increase the efficiency of the actual interfaces. To allow the experimentalist to feel and touch the nano-world, we add mixed realities between an AFM and a force feedback device, enriching thus the direct connection by a modeling engine. We present in this paper the first results from a real-time remote-control handling of an AFM by our Force Feedback Gestural Device through the example of the approach-retract curve.

## 1 Introduction

The development of Scanning Probe Microscopy (Scanning Tunneling Microscopy, Atomic Force Microscopy) allowed observing surfaces with a resolution, which can reach atomic scale. In addition, these instruments can be used as tools in order to modify surfaces [1] or to fabricate [2] and to manipulate objects and systems in such nano-scales. An AFM controls the coupling with the nano-world through the interaction of a nanometric tip with nano-objects or surfaces.
Depending on these types of interaction, the manipulation possibilities with an AFM can be: mechanical contact used to deform, press, push/pull, displace and assemble nano-objects [3] or electrostatic coupling used to deposit, move and detect electrons cloud on an insulating surface or charged nano-objects [4].
Nevertheless, in nowadays situation, with the basic commercial AFM interfaces, the user is restricted to blind movements because the operation of object displacement can be executed only with automated task without getting any feedback reaction loop, which can lead to sample or tip damage and which is time consuming. The interest of rendering a real time feedback control by connecting the user into a real-time interaction loop with the nanotool will avoid the trials and errors process, unwanted displacements and long scans to check the result of the action. By this way, the user will be able to use the AFM such as a musician would use his instrument, and physically interact with the sample, like with a human-scaled sample. In order to reach this goal, the actual tendencies in nanotechnology are pointed for creating advanced nano-manipulators [3]. They are requested in many applications [5, 6] from biology [7] to nanoelectronics, and as a direct and sensorial approach for the theoretical descriptions of the nano-world in educational sciences [8].

Adding mixed realities between an AFM and a Force Feedback Gestural Device (FFGD) increases the on-line efficiency during the nano-object observation and manipulation: the user can reach directly his goals thanks to computer-assistance and through multi-sensorial perceptions (haptic, visual and acoustic). Using this modeling engine, the user can also train to manipulate in a virtual nano-scene and evaluate the result of these actions.

This paper focuses on the use of an AFM in contact mode and more precisely on the force curve mode. A force curve measures the vertical force resulting from the interaction between the AFM tip and the sample in function of the tip-surface distance resulting in a hysteresis phenomenon. The presence [9] of this phenomenon in the user's space can be reconstructed only by combining and synchronizing the intensity, the place and the time of action in the multi-sensory rendering as it shows in this paper results. We will first give a more detailed description of the approach-retract phenomenon, and then we will describe the full 1-DOF nano-manipulation chain, using our custom-made FFGD. The final part will enhance the necessity of a specific force feedback loop between the user and its tool, and the way we can use the multisensory devices for rendering the best feedback to the user (sound, images and forces) and allow a real time effective interaction.

## 2  AFM force curve: a 1-DOF nano-scale phenomenon

The existing local force between the AFM tip and a surface in a cohesive interaction generates a complex and non-linear force-position characteristic curve [10]. The measurement technique consists in recording the cantilever deflection when the AFM probe approaches at constant speed in the vertical direction towards the surface, above a fixed point of the sample and then, retracting it. During this measurement the feedback loop of the microscope is turn off.

Different cantilever states (Fig.1) can be distinguished on the force curve:
- During the tip-surface approaching, A is the null deflection cantilever state corresponding to a free descending, B is the snapping state corresponding to the moment when the force gradient of the tip-surface interaction get greater than the stiffness of the cantilever, then C corresponds to the repulsive regime when the cantilever deflection increases with a constant slope $\chi = \alpha\kappa/(\alpha + \kappa)$ (where $\alpha$ is the repulsive slope of the tip-surface interaction and $\kappa$ is the stiffness of the cantilever). If the sample is rigid ($\alpha \gg \kappa$), $\chi$ is nearly equal to the stiffness of the cantilever.
- During the tip-surface retracting, D is the state when the cantilever deflection decreases with the same slope $\chi$, E is the state of sudden increasing of the cantilever deflection corresponding to the reverse snapping when the force gradient became smaller than its stiffness.

This mode is used to determine local physical and chemical properties of the surface [11], to manipulate molecules [12]. Nevertheless, since the information is sent after its completion, the user can not react in real time to the process. The connection of a FFGD to an AFM, would allow to feel in real time the intensity and the variation of the interaction force, to know instantly what happens in the nano-scene and to adapt the movement of the tip consequently: to continue the approach or to retract the tip, to change the vertical speed of the tip.

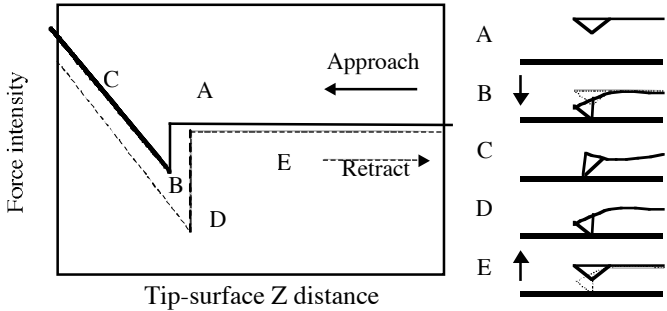

**Fig. 1.** Force curve according to tip-surface distance

In addition, thanks to this connection the user could choose more easily the intensity and the nature of the appropriate force to scan the surface in order to avoid the tip or surface damage.

## 3  Bidirectional teleoperation chain

The proposed instrumental chain consists in connecting the AFM to a multi-sensorial Real-Time Workstation equipped with a haptic device (Fig. 2), establishing thus a bidirectionnal link between nano and human world.
Due to the designed instrumental chain, an experimentalist can act on the haptic device key in order to command the AFM piezoelectric element in 1-DOF and thus the tip movement in vertical direction, while in reverse, the data transfer from the AFM to the haptic device, allows the experimentalist to feel in real time the forces existing between the AFM tip and the surface with a $10^8$ magnified factor. The used force feedback device is a home-developed manipulator [13] composed from independent bar keys. An electromagnetic actuator tracked by a high-resolution (2μm) position encoder controls these keys. A FFGD key is characterized by 20 mm vertical displacement, 10 kHz cutting frequency in force control loop, a maximum speed of 2 m/s, and supports a force of 50N in permanent regime and 200 N in transitory regime. In addition to this flexible morphology, the device power is greater than the up-to-now commercialized architectures where a force of 20N is hardly exceeded. The force applied on the key of the haptic device and its resulting position is the permanent information exchanged with the real-time workstation. Data transfer and computation take place at 3 kHz frequency in reactive mode, so the real-time condition is completely satisfied according to the human reaction time and the time constant of the simulated models [14].

Virtual models implemented on Real-Time Work-Station are based on the fundamental principles of physical simulation, using our custom-made Cordis-Anima language [15]. The desired components, modeled like material objects with specific weights and inertias, are permanently in interaction with the external environment as well as between them. Data concerning the position information received from haptic device and from AFM act on the virtual model which generates the resulting force data. The same force information is sent back to the haptic device as well as to the AFM piezoelectric element after the appropriate signal processing (Fig. 2). Using both positions and force data, a visual and sound rendering are created and offered to the experimentalist in addition to force rendering creating thus an interface that takes into account the full cognitive capabilities of the user.

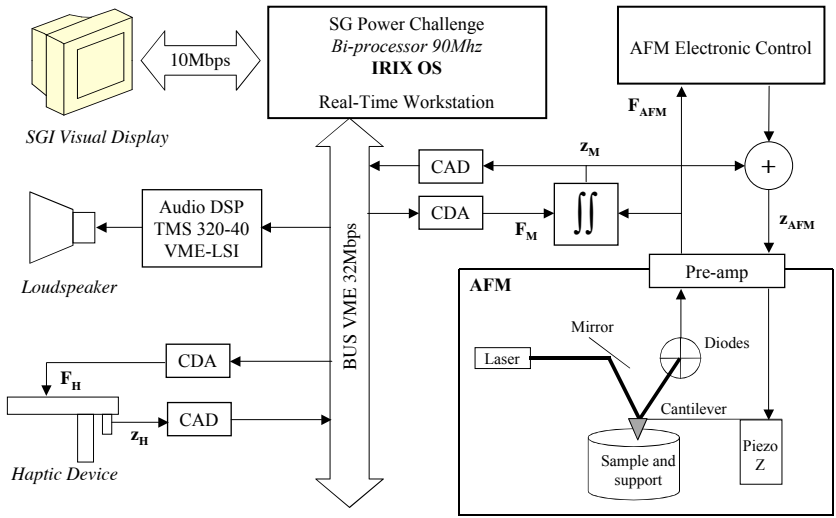

**Fig. 2.** Bidirectional teleoperation chain

To guarantee the sense of presence of the nano-objects that interacts: AFM tip and nano-surface, we carefully set the correspondence between the hand manipulation and the visual representation. Besides the reconstruction would not be complete without the auditory rendering. Adding the sound feedback to the haptic and visual sensory processes during action offers a complete multisensory environment.

The feeling of hysteresis described from the force curve is now brought back to our scale and the user hand has to thwart a dozen of newtons to reach the snap-off and release the tip of the AFM. In this work, the model simulated in the real-time workstation is a simple stiffness. The value of this stiffness is chosen in order to use the full capabilities of the force feedback system. This virtual spring is a basic link, which allows the dual transfer of forces and positions between the AFM and the FFGD. Thus this virtual component will surround and support not only the AFM-FFGD interconnection, but also the user gesture for a best control of the manipulation. In order to determine what kind of phenomena can be simulated in parallel with this interconnection, various virtual models connected only with the real-time multisensory loop are presented in the following.

# 4  Manipulation through virtual models connected to a multisensory architecture

Models presented in this part are implemented on the real-time workstation described above, using the Cordis-Anima language for physical simulation. They are connected to the FFGD through the same hardware architecture and thus are subjected to the same time constants. The AFM instrument does not belong to the loop any longer, and we model it in order to master its parameters and its behavior. As a validation, this modeling matches the behaviour of with the real AFM.

The virtual scene is based on a nano-surface model, subjected to elastic deformations due to the Lennard-Jones interaction between its elements and the virtual tip. A spring-modeled cantilever links the tip with the piezoelectric element. The user can move the virtual piezo element with the FFGD and feel the interaction force in a real-time loop. Besides the traditional AFM visual display, the rendering offers simultaneously a geometric representation of the scene, and on a phenomenon display: the evolution of the interaction potentials between elements, represented by a ball rolling down into the potential gap (Fig.3). The sound feedback strengths the multisensory rendering since it is directly captured from the acoustical vibration of the atomic layer, following the dynamics of the real time interaction.

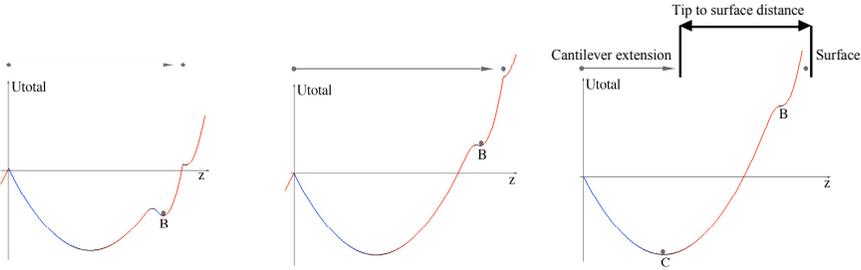

**Fig. 3.** Potential curves for visual representation during the retract movement

This representation is based on the addition (1) of the two potentials: the elastic potential characterizing the cantilever movement and the non-linear Lennard-Jones potential (2) characterizing the tip-surface interaction.

$$U_{total} = U_{cantilever} + U_{LJ} \quad (1)$$

Both visual representation and force feedback are based on the linearization of the force derivated from the Lennard-Jones potential. Linearization of Lennard-Jones potential is constituted of three parts.

$$U_{cantilever} = -\int F(z)dz = \lambda + K \cdot z^2/2, \quad U_{LJ} = \alpha/z^6 - \beta/z^{12} \quad (2)$$

Approaching the surface, the force feedback is null and no sound can be heard since there is no tip-surface interaction. Reaching the threshold, the tip suddenly snaps on the surface, the hand is pulled down by the FFGD and a wind-like sound starts to be heard. Simultaneously, the ball rolls down to a minimum hollow in the potential representation, trapped onto the surface in an equilibrium position.

The user can then decide to push on the surface in the same direction (Fig. 4a, extracted from the real-time visual rendering) and feel the stress due to the repulsive behavior of the interaction. The sound increases its harmonics while we see the visual deformation of the surface and feel the force feedback. Once the user try to pull the tip out of the surface, the sum of forces acting on the atomic layer of the surface become null, so the sound gradually disappears. If we continue pulling the AFM probe (Fig. 4b), the sound increases again according to the absolute force value of the attractive interaction. The geometric display shows that the cantilever is extended, and the ball in the potential representation is brought higher. Once the ball is about to fall in the new hollow (Fig. 3), an instable regime shifts the sound towards low frequencies. The snap-off brings the system back to the original state.

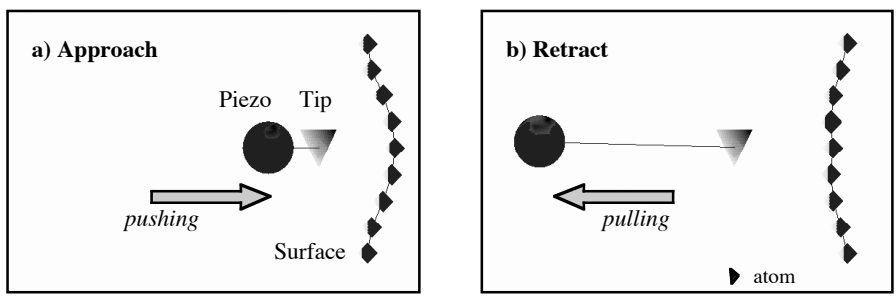

**Fig. 4.** Feeling the strain or the expansion of the atomic layer (surface)

This virtual model emphasizes the ability of the user to control the movement without wasting time parametering an automatic process (tip velocity, piezo elongation). We noticed that the learning time for our students to find the correct movement can be reached nearly from the first action, with a very high accuracy and reactivity, thus decreasing the experiment time by three or four. Due to the intuitive understanding of the phenomenon, the use of the natural dexterity of our actions and the transparent and energy-conserving behaviour of the system, the user will permanently adapt the impedance of his hand in order to prevent the tip to snap-off, according to his multisensory feelings.

## 5   Conclusion

The force feedback feeling given by our instrument appears much more relevant than the simple observation of an approach-retract curve, since the user can interact in real time with the surface, and thus perform a complex action: modulate the speed, stop the movement, approach or retract with a real time dynamic response, events which cannot be done without this specific architecture. Simulated parts introduced in the communication chain allow implementing additional intelligent support and present a more complex input/output set up. The introduction of a simulator between the manipulation instruments existing in user space and the manipulated nano-scene increases the online efficiency: the operator is able to interact with the virtual scene so as to study its properties and to rehearse the actions he plans to undertake, subsequently, on the real scene. This simulator plays as an intelligent emulator that assists the experimentalist and adds intelligent high-level processes in the control of the nanosystem, in the real phenomena reconstruction.